\documentclass[manuscript,screen]{acmart}

\usepackage{url}
\usepackage{makecell}
\AtBeginDocument{}

\setcopyright{none}

\usepackage{xspace}
\usepackage{cleveref}
\usepackage{listings}
\usepackage{subcaption}
\usepackage{xcolor}
\usepackage{siunitx}
\usepackage{textcomp}
\usepackage{balance}
\usepackage{tabularx}
\usepackage{booktabs}
\usepackage{tcolorbox}

\tcbuselibrary{listings,breakable,skins}

\newtcblisting{promptbox}[1][]{
  enhanced,
  breakable,
  listing only,
  colback=white,
  colframe=gray!65,
  colbacktitle=gray!65,
  coltitle=white,
  fonttitle=\bfseries\small,
  title={#1},
  boxrule=1.1pt,
  arc=3pt,
  boxsep=0pt,
  left=14pt,
  right=14pt,
  top=12pt,
  bottom=12pt,
  toptitle=6pt,
  bottomtitle=6pt,
  lefttitle=14pt,
  listing options={
    basicstyle=\ttfamily\footnotesize,
    breaklines=true,
    breakatwhitespace=true,
    columns=fullflexible,
    keepspaces=true,
    showstringspaces=false,
  },
}

\usepackage{pifont}
\definecolor{figink}{HTML}{    5E8734}
\newcommand{\cone}  {\textcolor{figink}{\ding{182}}}
\newcommand{\ctwo}  {\textcolor{figink}{\ding{183}}}
  
\newcommand{\eg}{e.g.,\xspace}
\newcommand{\ie}{i.e.,\xspace}

\newcommand{\tool}{HANSEL\xspace}
\newcommand{\fullname}{Highlighting Agent Navigation Steps as Evidence Links}

\renewcommand\footnotetextcopyrightpermission[1]{}

\begin{document}

\title{\tool: Interactive Evidence for Post-Hoc Verification of Web Agents}

\author{Yujin Zhang}
\email{yujinz16@uci.edu}
\affiliation{
  \institution{University of California, Irvine}
  \city{Irvine}
  \state{California}
  \country{USA}
}

\author{Yeonju Hwang}
\email{rhdwn1804@kookmin.ac.kr}
\affiliation{
  \department{Software Engineering}
  \institution{Kookmin University}
  \city{Seoul}
  \country{Republic of Korea}
}

\author{Daye Nam}
\email{daye.nam@uci.edu}
\affiliation{
  \institution{University of California, Irvine}
  \city{Irvine}
  \state{California}
  \country{USA}
}

\begin{abstract}
AI web agents can perform complex tasks, such as searching for products and making purchases on users' behalf. However, verifying an agent's output remains difficult, because existing transparency mechanisms (e.g., full trajectory logs) require users to review logs and explanations, keeping verification a passive reading task. 
In this work, we view agent verification as an interactive activity and present \tool.
Given an agent trajectory, \tool extracts evidence pages and snippets and presents them as navigable, live views that preserve relevant page state (e.g., applied filters), allowing users to verify how the agent arrived at its answer through interaction.
In a controlled user study with 14 participants, we show that \tool significantly reduces task completion time and perceived effort, demonstrating that reframing verification as an interactive activity enables more efficient post-hoc verification of AI agents.
\end{abstract}


\begin{teaserfigure}
  \centering
  \includegraphics[
  width=1.0\textwidth,
  trim=0 0cm 0 0.5cm,
  clip]{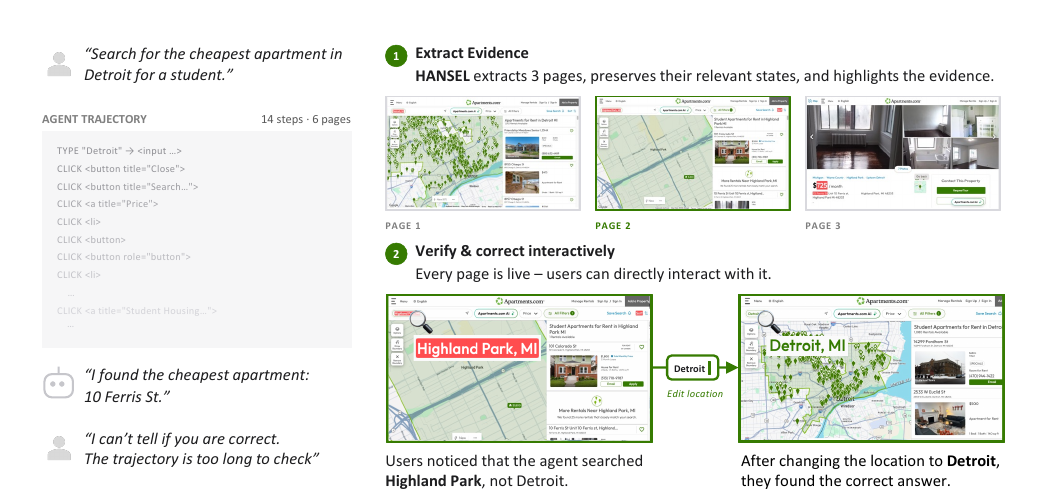}
  \caption{Verifying an agent's output often involves inspecting long and complex trajectories (\eg 14 steps, 6 pages visited).
  \cone~\tool extracts the important visited evidence pages and presents them as interactive pages with relevant page states preserved (\eg applied filters, search queries, and scroll positions). Relevant evidence is highlighted in red.
  \ctwo~Users can interact with these evidence pages to verify and correct the agent's answer if needed. For example, a user can identify that the agent searched the wrong location and directly change it on the evidence page to find the correct results.
  }
  \Description{}
  \label{fig:teaser}
\end{teaserfigure}

\maketitle

\section{Introduction}

AI agents are capable of performing multi-step, complex tasks on our behalf: agentic systems can now search for products, compare options across multiple platforms, and even make purchases or send messages on behalf of users~\cite{deng2023mind2webgeneralistagentweb, zhou2024webarenarealisticwebenvironment,mialon2023gaiabenchmarkgeneralai}.
As these agents take on tasks with non-trivial real consequences, a critical question emerges: how can users verify that the agent's answer is actually correct~\cite{bowman2022measuringprogressscalableoversight,bansal2024challengeshumanagentcommunication}?

Consider the following query posed to a web agent: \textit{``Search for the cheapest apartment in Detroit for a student on Apartments.com.''}
The agent must navigate Apartments.com, search for ``Detroit, MI,'' apply a student filter, sort prices from low to high, and return the result.
Many web agents can correctly find answers to queries like this.
However, when the agent returns ``10 Ferris St'', how likely is a user to contact the apartment manager for a viewing with full trust, without verifying whether the agent missed cheaper options within the area or made an error?
If the tasks had higher stakes, for example, making a purchase or sending an email to customers, the desire to understand what agents did would only increase.

Recent systems support user monitoring and intervention during agent execution through mechanisms such as co-planning and action confirmation~\cite{mozannar2025magenticuihumanintheloopagenticsystems,cowpilot2025huq, feng2025cocoacoplanningcoexecutionai}. 
However, requiring users to remain attentive throughout execution conflicts with the common usage scenario and motivations of agents, as users delegate tasks not because they lack the ability to perform them, but to offload tedious, repetitive, or time-consuming work~\cite{ning2025surveywebagentsnextgenerationai,lubars2019askaidoai}.
Rather than requiring continuous attention, systems should support \textit{post-hoc verification}, helping users verify after execution that the agent completed the task correctly~\cite{lim2009why, bansal2024challengeshumanagentcommunication}.

To verify an agent post-hoc, users need to understand how the agent reached its answer and judge whether its actions and the information it relied on are correct.
However, existing transparency mechanisms do not meet these requirements.
Execution logs expose lengthy traces that users must sift through.
Other mechanisms such as source links~\cite{gemini_support_2025}, screenshots~\cite{openai_operator_2025}, and reasoning summaries~\cite{nakano2022webgptbrowserassistedquestionansweringhuman} are more compact, but lose the interaction context or fidelity needed to verify what the agent actually did.
The gap common to all these mechanisms is that they treat verification as a passive reading task: users are shown what the agent did but receive little support for engaging with it directly.
This directly resembles the overreliance problem identified in AI-assisted decision-making, where simply presenting information such as explanations does not lead people to critically engage with AI outputs, and interaction designs that more actively involve users are needed~\cite{buccinca2021totrustortothink}.
Presenting static information is not enough; users need to actively examine and interact with the evidence behind an agent's answer.

We present \textbf{\tool} (\fullname), a system that supports 
interactive post-hoc verification.
Instead of requiring users to continuously monitor agent execution or reconstruct the task from scratch, 
\tool identifies the subset of visited pages that the agent relied on to produce the final answer and presents them as navigable, interactive views with relevant state preserved (\eg applied filters).
As illustrated in Figure~\ref{fig:teaser}, \tool extracts three evidence pages with relevant evidence highlighted. 
Users can identify that the agent searched the wrong location and correct it directly on the evidence page post-hoc, rather than restarting the task from scratch.

The design of \tool is guided by three goals.
First, \tool \textit{reduces evidence to what matters}. 
Rather than presenting the execution logs, which are too overwhelming, or a single visited page, which is too narrow, \tool surfaces the compact set of visited pages that the agent actually relied on to produce its answer, balancing completeness with cognitive load. 
Second, \tool \textit{makes evidence inspectable and gaps visible}; 
a URL alone is not sufficient if users land on a default page and must reconstruct the agent’s interactions to locate the relevant information. \tool instead preserves the page state the agent encountered (\eg scroll positions, applied filters, and search queries) and highlights relevant evidence, allowing users to quickly inspect what supports the answer and identify when such support is missing.
Finally, \tool \textit{lets users act on what they find}: when users need to further verify the evidence or notice a potential problem with the agent’s answer, they can interact directly with the evidence page rather than being limited to static screenshots or textual explanations. 
Together, these three goals define \textit{interactive verification}: turning an agent’s execution trace into interactive evidence that users can inspect in context and act on directly to verify the answer.

We evaluated \tool both with technical benchmarks and with users.
Our technical evaluation revealed that \tool accurately identifies core evidence pages from agent trajectories, achieving 80.2\% precision and 86.3\% recall against ground-truth annotations across 96 tasks from AssistantBench~\cite{yoran2024assistantbenchwebagentssolve} and Online-Mind2Web~\cite{xue2025illusionprogressassessingcurrent}.
In a controlled user study with 14 participants, \tool significantly reduced completion time and perceived effort among correctly completed tasks. 
Participants also rated \tool as useful and found its key features helpful.
These demonstrate the feasibility and the usefulness of interactive post-hoc verification of agent actions.

In summary, we contribute:
\begin{itemize}
    \item A formulation of the agent evidence extraction for human post-hoc verification problem,
    \item A human-annotated dataset of 96 real-world web agent trajectories, comprising 1,177 execution steps and 497 visited pages,
    \item An interactive verification interface that enables users to engage directly with evidence, rather than passively consuming agent explanations,
    \item Empirical evidence from a user study demonstrating that interactive evidence presentation leads to more efficient verification.
\end{itemize}

\section{Background \& Related Work}

\subsection{LLM-Based Web Agent}
With the advancement of LLM capabilities in web navigation~\cite{xu2025agenttrekagenttrajectorysynthesis, murty2025nnetnavunsupervisedlearningbrowser}, web navigation tasks have shifted from reinforcement learning (RL) tasks~\cite{gur2018learningnavigateweb, pmlr-v70-shi17a} toward LLM-based web agents~\cite{nakano2022webgptbrowserassistedquestionansweringhuman, he2024webvoyagerbuildingendtoendweb, zheng2024gpt4visiongeneralistwebagent}.
These agents show strong performance on web agent benchmarks spanning information seeking, website interaction, and long-horizon task completion~\cite{deng2023mind2webgeneralistagentweb, zhou2024webarenarealisticwebenvironment, yao2023webshopscalablerealworldweb, jang2026odysseysbenchmarkingwebagents}.
However, web agents still face challenges in real-world web tasks.
Agents frequently make grounding errors, failing to correctly identify and interact with the intended page elements~\cite{deng2023mind2webgeneralistagentweb, yao2023webshopscalablerealworldweb}.
They also get stuck in loops and fail to recover in long-horizon tasks~\cite{yoran2024assistantbenchwebagentssolve, chung2025evaluatinglongcontextreasoningllmbased}.
Even when agents correctly understand the task and perform reasonable actions, they often miss crucial details and lose track of the original task objectives~\cite{lu2025agentrewardbench,aghzal2026llmbasedwebagentsfail}.
Real-world websites introduce additional challenges, such as unpredictable bot detection mechanisms, which can break automation and require human intervention~\cite{xue2025illusionprogressassessingcurrent}.
Despite these failures, agents can still produce a result even when errors are hidden in intermediate steps and are not visible in the final response, motivating us to design \tool to help users do post-hoc verification and correction of agent actions.

\subsection{AI Agent Monitoring}

Prior HCI research has examined how people verify AI-generated outputs and assess their correctness~\cite{gu2024analystsunderstandverifyaiassisted,relic2024cheng, zhou2026improvinghumanverificationllm}.
For AI agents, recent work has explored approaches that allow users to monitor and intervene in agent execution in real time~\cite{cowpilot2025huq,mozannar2025magenticuihumanintheloopagenticsystems, Morae2025Peng, feng2025cocoacoplanningcoexecutionai}.
Magentic-UI, for example, provides a live browser view and allows users to steer, approve, or directly take over agent actions~\cite{mozannar2025magenticuihumanintheloopagenticsystems}.
Morae proactively detects ambiguous decision points and pauses UI automation for users to steer~\cite{Morae2025Peng}.
Similarly, Zhou et al. model the placement of intermediate confirmation points in multi-step agent tasks to decide when to stop for users to check~\cite{Zhou2026when}.
However, these approaches primarily support verification during execution, requiring users to remain involved as the agent works. 
Indeed, such involvement may not always align with how agents are used in practice, where users delegate tasks to agents to reduce their own effort and attention~\cite{lubars2019askaidoai,dhanorkar2026humanoversightagenticsystems}.

\subsection{Post-hoc Verification Mechanisms}
\label{subsec:why_current_mechanisms_fall_short}

While sparse, there has been some research done to create post-hoc verification support, and modern web agents also offer several forms of transparency to help users understand how the agent produced the answer.
However, most of the current mechanisms fall short, and we discuss why they may not be sufficient for users to grasp what the agent actually did.

\subsubsection{Reasoning Summaries}
The most well-studied and used mechanism in conveying the agent's reasoning is providing a natural language summary.
Many agents such as WebGPT~\cite{nakano2022webgptbrowserassistedquestionansweringhuman} produce natural-language summaries of their process.
Some systems provide static snapshots of pages that the agent visited as part of the summary~\cite{mozannar2025magenticuihumanintheloopagenticsystems, openai_operator_2025}.
However, LLM-generated summaries and explanations may not faithfully reflect the process underlying their behavior~\cite{turpin2023faithful, lanham2023measuringfaithfulnesschainofthoughtreasoning, Andreas2024faithful}, and such unfaithfulness has been observed even in highly capable models~\cite{tanneru2024hardnessfaithfulchainofthoughtreasoning, bentham2024chainofthought}.
For example, a summary may describe how the agent successfully identified a product satisfying all constraints, while the agent in fact failed to apply the correct filters and missed a qualifying option.
Users can only verify the summary by returning to the agent’s actual execution.
Moreover, plausible-sounding explanations may reduce users’ cognitive engagement and encourage overreliance~\cite{Gajos_2022}.

\subsubsection{Source Page Links}
Another common mechanism, especially for commercial tools, is to provide source links to webpages the agent has visited for users to review~\cite{gemini_support_2025, openai_chatgpt_search_2026}.
For example, Gemini provides source links either inline or through a side panel, allowing users to access related webpages for further inspection~\cite{gemini_support_2025}.
However, when source webpages and execution traces are presented separately, users must locate the relevant information and reconstruct its connection to the agent’s actions themselves.
For complex web tasks, source links may also fail to preserve important interaction context, such as the webpage state encountered by the agent.
For example, if the agent sorted results by price and filtered by location, users who follow the link may see the page without these settings and must manually reconstruct the agent’s search context to verify the answer.

\subsubsection{Execution Logs}
The richest and most extensive information to provide to the user is the execution trajectories of the agent.
Systems built on ReAct-style reasoning~\cite{yao2023react} provide the agent's complete action-and-reasoning trace.
While comprehensive, these logs can be overwhelming: all steps, including navigation overhead, abandoned search paths, and failed interactions, are displayed without curation, making it difficult for users to identify what to focus on.
For example, in a shopping task, the agent may repeatedly refine and retry searches while switching between listing and detail pages without making progress, with each attempt logged with detailed actions (\eg \texttt{type}, \texttt{click}, \texttt{scroll}) and reasoning traces (\eg ``Failed'', ``I need to ...''), even though none of them contribute to the final answer.

Some recent work has developed interfaces for organizing and visualizing agent trajectories, such as structured step-level views~\cite{openhands_trajectory_visualizer_2025}, diagnostic annotations~\cite{ou-etal-2025-agentdiagnose}, overview panels~\cite{interactive2025Epperson}, and flowcharts~\cite{wang2026xagenexplainabilitytoolidentifying}, which could help with better understanding.
However, most of them are built for AI tool builders for debugging and failure analysis, rather than for end-users. Therefore, the execution-level details can still be overwhelming for end-users and may not provide an intuitive way to understand the agent’s work.

\subsubsection{The Gap}
Across these mechanisms, the main gap is that verification is treated primarily as a passive reading task: users must consume logs, screenshots, links, or textual explanations.
However, simply providing such information does not necessarily lead to critical engagement~\cite{buccinca2021totrustortothink}, while providing additional explanations can increase users' time cost without improving their decision accuracy~\cite{chen2026aiknowsbest}.

\section{Formative Analysis and Design Goals}

In this section, we further understand and quantify the gap we identified in the previous section, and synthesize the properties that effective post-hoc verification requires.
In the formative study, we specifically focus on execution logs and source page links, whose usefulness is less understood compared to reasoning summaries~\cite{chen2026aiknowsbest, buccinca2021totrustortothink}.

\subsection{Empirical Analysis of Agent Trajectories}
\label{subsec:empirical_analysis}

To quantify the challenges of using execution logs and source page links for post-hoc verification, we manually analyzed agent trajectories for web agent tasks.
We selected two popular web agent benchmarks, 
AssistantBench~\cite{yoran2024assistantbenchwebagentssolve} and Online-Mind2Web~\cite{xue2025illusionprogressassessingcurrent}.
AssistantBench is a benchmark of realistic, time-consuming information seeking tasks on the open web, each with a ground-truth answer.
It provides a development set with 33 tasks for experimentation and analysis.
Online-Mind2Web~\cite{xue2025illusionprogressassessingcurrent} is a benchmark of 300 realistic web navigation tasks spanning 136 websites, covering a broad range of procedural goals.
We collected the actual agent trajectory logs from HAL~\cite{hal2025}, which evaluates AssistantBench and Online-Mind2Web using browser-use~\cite{browseruse_github}, a Python-based Playwright browser automation agent.
We selected the top-performing model for each benchmark at the time of our study based on the reported accuracy in HAL.
For AssistantBench, we used trajectories generated with o3-medium (38.8\% accuracy) in HAL.
We removed tasks where the agent failed to produce an answer, and those where no web interaction was performed (i.e., the agent directly produced an answer in the first step without any web actions), resulting in a final set of 22 tasks.
For Online-Mind2Web (300 tasks), we used trajectories generated with Claude Sonnet 4 (40.0\% accuracy).
We excluded open-ended tasks that did not have a single verifiable ground-truth answer (\eg \textit{``Compare Apple Watches and learn more about the Ultra version on Apple''}) and kept 74 tasks.
This resulted in a combined analysis set of 96 tasks (22 from AssistantBench and 74 from Online-Mind2Web).

For each task, two researchers independently examined the full trajectory and labeled each step as an evidence step or a non-evidence step.
The annotation process took approximately 20 hours in total.
We annotated the trajectories at two levels: steps and pages.
A step is an atomic observation-reasoning-action unit, consisting of an observation of the current page state, reasoning about the next goal or plan, and action execution, followed by immediate environmental feedback.
A step was labeled as an evidence step if its content or visible page state was needed to trace or verify how the agent arrived at its final answer.
This includes process evidence, such as search queries, selected filters, sorting criteria, and result pages, as well as content evidence, such as facts, values, or constraints used in producing the answer.
Steps corresponding to failed attempts, navigation overhead (\eg loading homepages, scrolling), blocked pages (\eg login walls, bot protection), or repeated retries without new information were excluded. 
A page is defined as a unique URL visit. 
Consecutive steps sharing the same URL are grouped into a single page, while non-consecutive visits to the same URL are counted as separate pages. 
For example, in a trajectory \texttt{page1} $\rightarrow$ \texttt{page2} $\rightarrow$ \texttt{page3} $\rightarrow$ \texttt{page1}, the two \texttt{page1} visits are counted as two separate pages. 
A page was labeled as an evidence page if it contained at least one evidence step.
Inter-rater agreement reached Cohen’s $\kappa$ of 0.803 at the step level and 0.878 at the page level, indicating good agreement.
Disagreements were resolved through discussion between the two annotators to produce the final labels.

\begin{table}[h]
\centering
\caption{Trajectory and evidence statistics for the 96-task analysis dataset.}
\label{tab:trajectory_analysis}
\begin{tabular}{>{\raggedright\arraybackslash}p{2cm}ccc}
\toprule
 & AssistantBench & Online-Mind2Web & Overall \\
\midrule
Tasks                             & 22             & 74             & 96             \\
Total steps                       & 204            & 973            & 1,177          \\
Total pages                       & 146            & 351            & 497            \\ \addlinespace[1ex]
Evidence steps                    & \makecell[t]{44 \\ (21.57\%)}   & \makecell[t]{253 \\ (26.00\%)}  & \makecell[t]{297 \\ (25.23\%)}  \\ \addlinespace[0.5ex]
Evidence pages                    & \makecell[t]{36 \\ (24.66\%)}   & \makecell[t]{175 \\ (49.86\%)}  & \makecell[t]{211 \\ (42.45\%)}  \\
\bottomrule
\end{tabular}
\end{table}

Across the 96 trajectories, agents executed an average of 12.26 steps and visited 5.18 web pages per task (Table~\ref{tab:trajectory_analysis}), yet only 3.09 steps (25.23\%) and 2.20 web pages (42.45\%) on average were identified as evidence.
The remaining steps corresponded to navigation overhead (\eg loading homepages, scrolling), abandoned search paths, or failed interactions.
Although Online-Mind2Web had longer trajectories than AssistantBench (13.15 vs.\ 9.27 steps per task), the fraction of evidence steps remained small in both datasets (26.00\% and 21.57\%, respectively).
The page-level evidence ratio in Online-Mind2Web appears higher because many within-page actions are irrelevant to the final answer, even when they occur on an evidence page.
This analysis empirically validates that the small subset of steps that matter for verification is often buried within a much larger volume of irrelevant information.

\subsection{Design Goals}

Based on the limitations of current post-hoc verification mechanisms and the formative analysis of the actual agent trajectories, we synthesize the following design goals for a post-hoc verification tool for AI agent users. The tool should be able to:
\begin{itemize}
    \item \textbf{DG.1: Surface a Compact Evidence Set:} 
    such that, users can verify the agent's answer more quickly with a subset of evidence pages that capture how the agent arrived at the answer.
    \item \textbf{DG.2: Enable Direct Inspection to Find Errors:}
    such that, users can directly inspect the evidence through different views of the
    interactive pages reconstructed by \tool with critical state preserved. If
    the agent fails, users can notice where it went wrong.
    \item \textbf{DG.3: Support Immediate Repair in Context:}
    such that, users can revise the answer or resume the task on the very page where the
    problem surfaced, instead of rerunning the task from scratch by themselves.
    
\end{itemize}

\section{\tool}

\begin{figure}
    \centering
    \includegraphics[trim={1.5cm 3cm 1.5cm 2cm},clip,width=\linewidth]{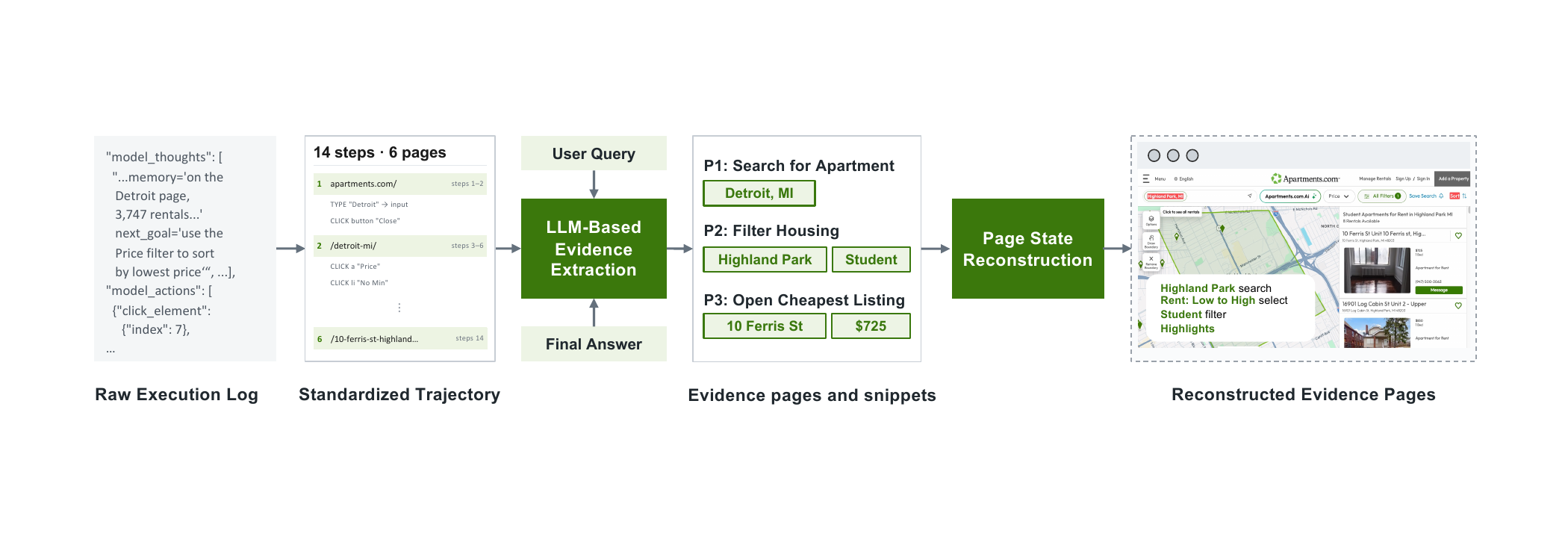}
    \caption{Overview of the \tool pipeline. \tool standardizes the raw agent execution log, uses an LLM to extract evidence pages and snippets, and reconstructs the identified pages with evidence highlighted in red.}
    \label{fig:hansel_overview}
\end{figure}

Based on our design goals, we designed and implemented \tool to support \textit{efficient post-hoc agent verification using interactive evidence}.

\subsection{DG.1: Surface a Compact Evidence Set}
\label{subsec:dg1}

In our empirical analysis of agent trajectories (Section~\ref{subsec:empirical_analysis}), only a small fraction of pages were directly used to produce the final answer, suggesting that surfacing this subset of evidence pages can make verification more efficient by reducing the information users need to inspect.
Based on this finding, we develop an evidence extraction pipeline that automatically identifies \emph{evidence pages} from an agent's trajectory.
We further extract \emph{evidence snippets}: textual content and localized interactive page states within an evidence page that the user needs to attend to.
We extract this more granular information because a single page may contain multiple steps, only some of which are relevant.
For example, within a single evidence page, the agent may type in search boxes to get autocomplete suggestions, scroll to discover additional content, apply filters to narrow down options, and change sorting criteria.
If only the sorting criteria informed the final answer, the evidence snippet would consist of only the content related to the sorting criteria.
We now introduce the two components of this extraction pipeline.

\subsubsection{Execution Log Standardization.}
\label{subsubsec:standardization}
Raw execution logs contain rich information about what the agent did, including the full sequence of agent actions, intermediate states, and environment responses, so we treat them as the input to \tool.
However, raw logs are verbose, with much of the execution metadata irrelevant to evidence extraction, and heterogeneous across frameworks.
For example, browser-use scaffold logs a tree structure of LangChain operations with entries $\langle \text{id}, \text{parent\_id}, \text{op\_name}, t_{\text{start}}, t_{\text{end}} \rangle$.
In contrast, the HAL Generalist Agent scaffold organizes reasoning steps as a linear sequence $[(\text{role}_1, \text{content}_1), \ldots]$, with no analogous tree structure.
Despite incompatible architectures, both scaffolds implement the same observation-reasoning-action cycle.
We therefore normalize them into a single trajectory representation.
We represent an agent trajectory as an ordered sequence of $n$ interaction steps produced while solving a task, where each step records the agent's observation, reasoning, action, and the URL at which the action was executed:
\[
T \;=\; \langle s_1, \dots, s_n \rangle,
\qquad
s_i \;=\; (o_i,\; r_i,\; a_i,\; u_i).
\]

The observation $o_i$ captures the agent's environmental input at step $i$, typically a representation of the current page state (\eg raw HTML, a DOM snapshot, an AXTree view), and may also include the result of the previous action (\eg an error message or confirmation).
For example, one trajectory records the following page snapshot as an observation: \texttt{[0]<a title="" />}, \texttt{[1]<div />}, \texttt{[2]<textarea ...>}.
The reasoning $r_i$ describes the agent's thought about the current state, including how it evaluates the previous action results, interprets the observation, and decides the next action.
For example, when a CAPTCHA blocks access, the agent may reason that the previous goal failed due to the CAPTCHA page and decide to switch to an alternative search path.
The action $a_i$ is one of a set of predefined browser operations supported by the scaffold, such as \texttt{click}, \texttt{type}, and \texttt{search}, each executed with explicit parameters. For example, a click action specifies a target element index (\eg \texttt{click(index=24)}), while a search action specifies a query string (\eg \texttt{search(query="Detroit apartments")}).
The URL $u_i$ records the page on which $a_i$ was executed; it carries no information about the agent's reasoning, but it helps group steps into pages.

This standardized representation abstracts low-level execution details specific to each scaffold (\eg LangChain operation trees) while preserving the information necessary for evidence extraction and answer verification.

\subsubsection{LLM-Based Evidence Extraction.}

Following Section~\ref{subsec:empirical_analysis}, we group the steps of $T$ into
\emph{pages} $P(T) = \langle p_1, \dots, p_m \rangle$, where each page is a maximal run
of consecutive steps sharing the same URL.
The first goal of \tool is then to identify the evidence pages
$E(T) \subseteq P(T)$ and, for each evidence page $p \in E(T)$, the set of evidence
snippets $S(p)$ that capture information relevant to the agent's answer.
Identifying evidence requires determining which observed information contributed to the final answer rather than simply detecting which pages the agent visited.
We therefore use an LLM to extract the predicted evidence pages $\hat{E}(T)$ and, for each extracted page $p \in \hat{E}(T)$, the predicted evidence snippets $\hat{S}(p)$.
Specifically, we provide the LLM with: (1) the user query $q$, (2) the final answer $\hat{y}$, and (3) the standardized trajectory $T$ segmented into the pages $P(T)$.
We instruct the LLM to select the evidence pages from $P(T)$ and output the extracted evidence pages in the order they were visited.
For each page, the output contains a brief title and a more detailed description of the agent’s actions on this page, as well as all evidence snippets identified on that page.

To ensure that the extracted evidence is concise and relevant, we design a set of prompt rules.
Since agents often follow multiple different plans given a single task, or revise their plans during execution if their original plan fails, we instruct the LLM to discard abandoned plans, and report the reasoning path that led to the agent’s reported answer.
We also instruct the LLM to remove blocked pages (\eg connection errors, login gating, bot protection), and intermediate pages resulting from redirects, as these do not provide usable information for answering the task.

\subsection{DG.2: Enable Direct Inspection to Find Errors}

\begin{figure*}
    \centering
    \begin{subfigure}[t]{0.48\linewidth}
        \centering
        \includegraphics[trim={0 0.5cm 0 1.0cm},clip,width=\linewidth]{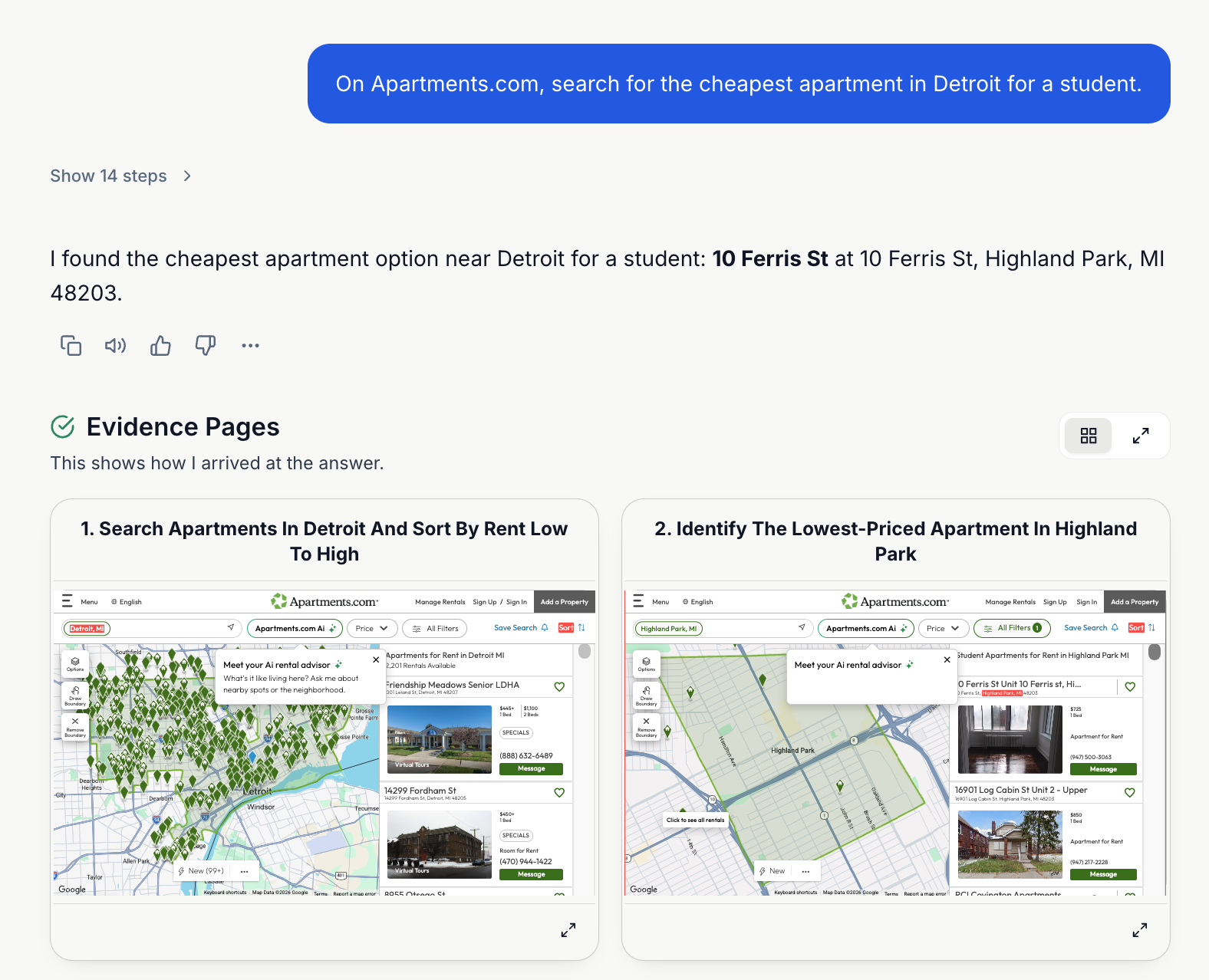}
        \caption{Grid View in \tool}
        \label{fig1:grid}
    \end{subfigure}
    \hfill
    \begin{subfigure}[t]{0.48\linewidth}
        \centering
        \includegraphics[trim={0 0.3cm 0 0.8cm},clip,width=\linewidth]{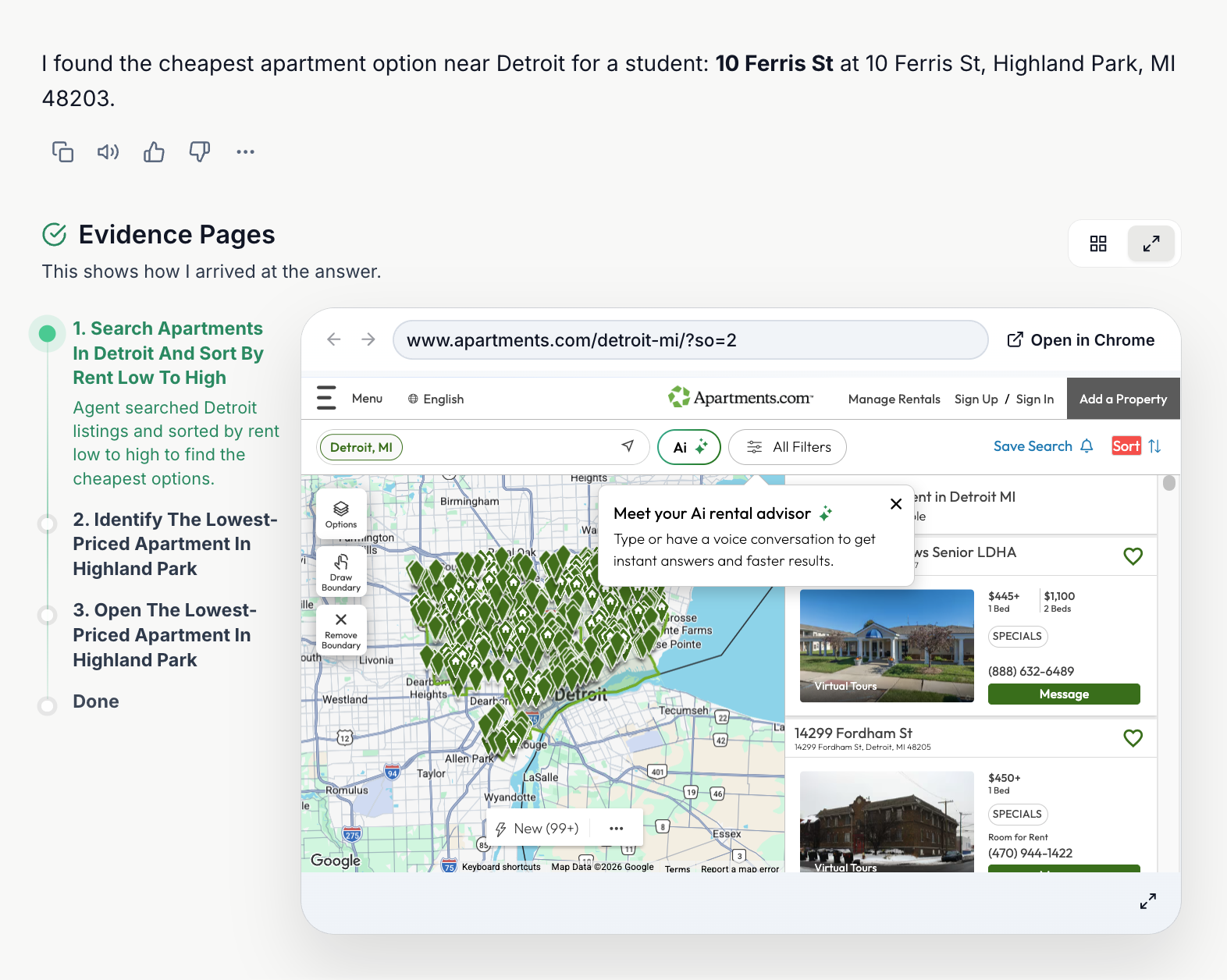}
        \caption{Carousel View in \tool}
        \label{fig1:carousel}
    \end{subfigure}
    \caption{\tool provides two complementary layouts for inspecting interactive evidence pages. (a) The grid view presents multiple evidence pages simultaneously for a high-level overview. (b) The carousel view presents evidence pages sequentially, with a timeline for navigation and a larger reconstructed webpage for detailed inspection and interaction.}
    \Description{}
    \label{fig:dg2}
\end{figure*}

To help users inspect the evidence and recognize when the agent's claims may be incorrect, \tool offers multiple ways of displaying evidence pages and their snippets (Figure~\ref{fig:dg2}).

\tool presents evidence pages in two complementary layouts: grid view (Figure~\ref{fig1:grid}) and carousel view (Figure~\ref{fig1:carousel}). 
In both layouts, we provide a button that allows users to expand an individual evidence page within \tool for closer inspection and interaction with highlighted evidence (expansion view), or open it in an external browser.

The grid view displays multiple evidence pages simultaneously, arranged in order based on their timestamp in the agent’s trajectory.
This view is designed to help users easily spot errors in the high-level reasoning.
For example, if the agent simply provided an answer to the task based on its knowledge without visiting any supporting webpage, there will be no evidence page or snippet to show. 
In this view, each evidence page is associated with a concise, descriptive title that summarizes the agent’s action on that page, enabling users to quickly understand what the agent did.
In Figure~\ref{fig1:grid}, the evidence pages are presented. 
Users can expand any page to inspect it in detail and interact with it directly.

The carousel view organizes evidence pages based on their timestamps and presents them one by one, allowing users to inspect pages in order.
In the left panel, \tool provides a timeline with a short title and a detailed description for each page, allowing users to navigate between evidence pages.
In the right panel, \tool embeds a larger browser window that displays the reconstructed page with highlighted evidence, allowing users to inspect and interact with it directly.
Users can switch between these layouts depending on whether they want a global overview or a more detailed, step-by-step inspection.

On every page, \tool highlights the extracted evidence snippets directly on the live page.
For example, on the first evidence page in Figure~\ref{fig1:grid}, the apartment search location (\texttt{Detroit, MI} in the top-left corner) and the sort filter (\texttt{Sort}) are highlighted in red.
This helps users go straight to the snippets relevant to the agent's claims and
actions without scanning an entire page to find them.

\subsection{DG.3: Support Immediate Repair in Context}

Prior work has shown that agents can fail due to small interaction mismatches (\eg an incorrect sort option) rather than major reasoning errors~\cite{deng2023mind2webgeneralistagentweb, zheng2024gpt4visiongeneralistwebagent}.
These errors can be straightforward to fix, but they become costly when users must reopen pages externally and repeatedly switch back to the task interface to check. 
This motivates our design to render each evidence page as a live, interactive webpage embedded directly within \tool, where users can easily inspect and correct errors when necessary.
Users can directly interact with the embedded page, including clicking, scrolling, typing, or navigating, without leaving the \tool interface.
Each evidence page is initialized in a reconstructed state that mirrors what the agent encountered (\eg with filters applied or search queries pre-filled).
In the example shown in Figure~\ref{fig:teaser}, the agent searched for apartments in \texttt{Highland Park, MI} instead of \texttt{Detroit, MI}, but had correctly applied the \texttt{Student} filter and sorted the results by \texttt{Price (Low to High)}. Because the page is live and the rest of its state is preserved, users can simply update the location within \tool, without opening a browser, going to Apartments.com, searching \texttt{Detroit, MI}, and setting the filters.

\subsection{System Implementation}

For the evidence extraction, we use a Python-based pipeline.
\tool first converts the raw trajectory log into a standardized sequence of steps, where each step contains the agent’s observation, reasoning, action, and current URL. 
Given the user query, the final answer, and the standardized trajectory steps, \tool uses an LLM to identify visited pages directly relevant to the final answer and specific content on those pages that should be highlighted as evidence. 
We use \texttt{GPT-5.4} with a temperature of \texttt{0.7}.
The extracted results are stored as structured JSON objects, including page-level summaries and snippet-level evidence spans. 
The pipeline also parses the trajectory to recover the actions performed on each evidence page (\eg scroll, input, click), which are later replayed in the interface to reconstruct each page to the state the agent reached during task execution.
For the evaluation and user study, we implemented \tool to operate on existing agent trajectories collected from HAL~\cite{hal2025}.
The same pipeline can be applied to a new trajectory once an agent completes its execution.

The interface of \tool is implemented as a desktop application using React and Electron~\cite{electron}.
To enable users to view highlighted evidence and interact with live webpages in \tool, we initially explored using iframes to embed evidence pages.
However, iframes are often blocked by websites due to security restrictions.
We instead use Electron's BrowserView to embed a live, interactive browser instance, and \tool replays scripted actions such as waiting, clicking, scrolling, and selecting filters to reconstruct each page to the state the agent reached.
\tool then highlights the evidence snippets identified by the LLM directly on the page by injecting CSS styles into the embedded browser at render time.

\section{Evidence Extraction Evaluation}

Before testing \tool with human participants, we first tested whether the core evidence extraction component of \tool provides accurate evidence.
We used the agent trajectory dataset containing 96 tasks from AssistantBench~\cite{yoran2024assistantbenchwebagentssolve} and Online-Mind2Web~\cite{xue2025illusionprogressassessingcurrent},  after excluding tasks where the agent did not produce an answer and open-ended tasks without a single ground-truth answer; see details in \Cref{subsec:empirical_analysis}.
We evaluated extraction quality both at the page and snippet levels.
Table~\ref{tab:evaluation_results} summarizes the results.

\begin{table}
\caption{Page-level evidence extraction quality and snippet-level precision across two benchmarks.}
\centering
\small
\setlength{\tabcolsep}{4pt}
\begin{tabular}{lccccc}
\toprule
 & \multicolumn{4}{c}{Page-level} & Snippet-level \\
Dataset & \#Tasks & Precision & Recall & F1 & Precision \\
\midrule
AssistantBench  & 22 & 0.821 & 0.889 & 0.853 & 0.788 \\
Online-Mind2Web & 74 & 0.798 & 0.857 & 0.826 & 0.844 \\
\midrule
Overall         & 96 & 0.802 & 0.863 & 0.831 & 0.838 \\
\bottomrule
\end{tabular}
\label{tab:evaluation_results}
\end{table}

\subsection{Page-Level Extraction}
For page-level evidence extraction evaluation, we reused the annotated dataset from \Cref{subsec:empirical_analysis}.
Given each task trajectory $T$, we compared the extracted evidence-page set $\hat{E}(T)$ with the manually annotated ground-truth set $E(T)$.
We then computed Precision, Recall, and F1 from these aggregated counts (Table~\ref{tab:evaluation_results}).
Across 96 tasks, our method achieved an overall F1 score of 0.831 for identifying evidence pages relevant to the final answer.
\tool reduced the information volume from 497 trajectory pages to 227 extracted evidence pages (54.3\% reduction).

\subsection{Snippet-Level Precision} 
For the snippet-level precision evaluation, two annotators independently reviewed all 532 extracted evidence snippets for validity.
An evidence page can contain multiple evidence snippets (\eg filters and facts), and an evidence snippet was labeled as valid only if it was visually observable on the webpage and relevant to the agent's answer.
The annotators agreed on 484 of the 532 evidence snippets (90.98\% raw agreement), with a Cohen’s $\kappa$ of 0.63, indicating substantial agreement~\cite{landis1977measurement}.
Disagreements were resolved through discussion to produce the final labels.
Of the 532 evidence snippets extracted by the LLM, 41 out of 52 were judged valid on AssistantBench (0.788), and 405 out of 480 were judged valid on Online-Mind2Web (0.844). The overall snippet-level precision was 0.838 (446/532).

\subsection{Discussion}
Overall, \tool identifies evidence from agent trajectories with high accuracy, achieving over 80\% precision for both page-level and snippet-level extraction.
It also substantially reduces the amount of information presented to users, surfacing 2.36 pages on average out of 5.18 pages visited in the original trajectory.
We observed two types of errors.
First, the LLM sometimes selected pages that were related to the task but did not directly contribute to the final answer under our annotation protocol.
Second, the LLM occasionally mistook its own reasoning for page content, producing snippets that were not visually observable on the page.
Both failure modes help explain the observed precision errors, as they resulted in irrelevant or invalid evidence being included in the extracted evidence set.
We therefore proceeded to evaluate \tool with actual users.

\section{User Evaluation}

To evaluate whether making agent verification an interactive activity leads to more efficient post-hoc verification, we conducted a user study with 14 participants comparing \tool with a baseline interface.
Through this study, we answered the following research questions:
\begin{itemize}
    \item RQ1: Does \tool reduce the time and effort required to verify web agents' answers?
    \item RQ2: Does \tool influence users' ability to correctly verify web agents' answers?
    \item RQ3: How do users perceive and use \tool when verifying web agents' answers?
\end{itemize}

In this study, we used a within-subjects design with two conditions: (A) \tool, which provided interactive, evidence-driven post-hoc verification support; and (B) the baseline, a standard conversational agent interface that displays the agent's final response and visited source links. 
In both conditions, we showed a step-by-step action trajectory.
The study procedure and participant recruitment complied with our institution’s IRB requirements.

\subsection{Participants}

We recruited participants from a large research university through social media and university forums.
Participants were pre-screened to ensure they had used AI tools and were proficient in English.
Out of 14 participants, 6 identified as male and 8 identified as female.
Participants came from diverse academic backgrounds.
Seven participants were from computing and engineering fields, 3 from business and economics, 2 from law, 1 from cognitive science, and 1 from communication and journalism.
In terms of their familiarity with AI tools, all participants mentioned that they had used conversational AI tools, but eight participants (57\%) reported that they did not use the agent mode.
Two participants (14\%) reported only modest proficiency in completing web-based tasks.

\subsection{Tasks}

\begin{figure*}
    \centering
    \includegraphics[
      width=1.0\textwidth,
      trim=0 5.5cm 0 2cm,
      clip]{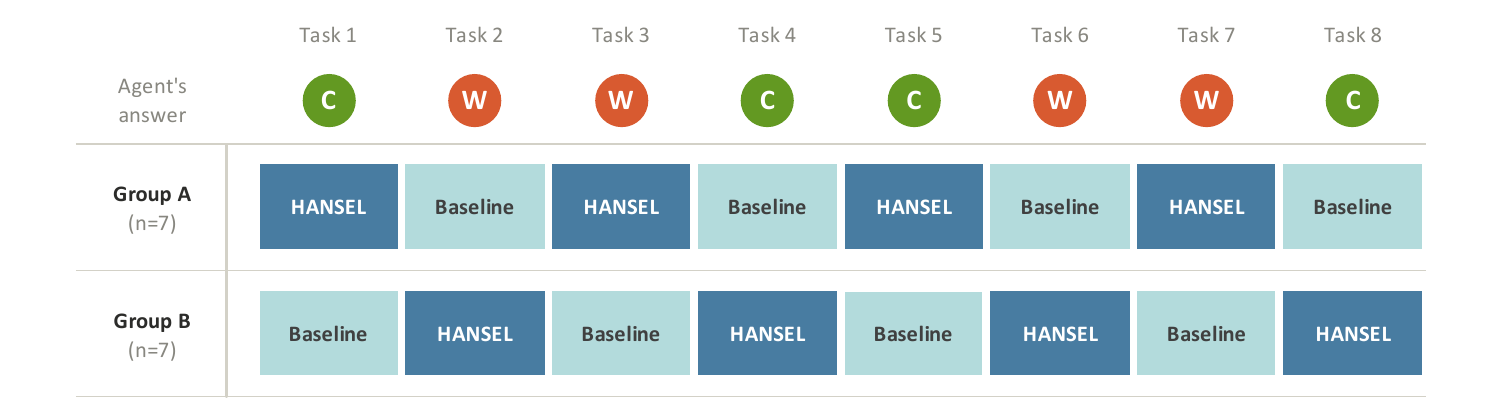}
    \caption{User study design across two groups. Participants alternated between \tool and the baseline across eight tasks, with agent answers following a fixed correct/wrong sequence. Tasks were paired by type: pricing calculation (Tasks 1 \& 4), sorting/filtering (Tasks 2 \& 7), selection under constraints (Tasks 3 \& 6), and information retrieval (Tasks 5 \& 8).}
    \Description{A two-row study schedule showing how participant groups alternated between HANSEL and the baseline across eight paired tasks while receiving a fixed sequence of correct and incorrect agent answers.}
    \label{fig:experiment_design}
\end{figure*}

We selected eight tasks to simulate scenarios in which users interact with web agents to assist with everyday tasks, from the manually annotated data in Section~\ref{subsec:empirical_analysis}.
We used the same agent trajectory for each task so that all participants evaluated the same agent execution and result.
We selected two tasks from AssistantBench and six from Online-Mind2Web.
The selected tasks involve multi-step information foraging, such as \textit{``How much would I save by getting an annual membership for my family (2 adults, 1 kid age 5, 1 kid age 2) living in Washington State for the Seattle Children’s Museum, compared to buying daily tickets, if we visit 4 times in a year?''}
We selected two tasks for each of four task types (\eg pricing calculation, information retrieval, sorting/filtering, and selection under constraints), so that one task of each type could be assigned to the baseline condition and the other to the \tool condition.

To test the interface's impact on participants' verification of agent responses, we selected four tasks in which agents returned correct answers and four in which they returned wrong answers.
Among the four agent responses with errors, two involved incorrect filtering or sorting (\eg failing to select the post with the most replies or the latest post), and two involved incorrect judgments based on webpage content (\eg selecting a product that was not the lowest priced, or searching in the wrong location, such as Highland Park instead of Detroit).
During the study, participants were assigned to use the two conditions alternately across eight tasks (Figure~\ref{fig:experiment_design}).
To minimize the noise coming from the order of the tasks, we kept the task order fixed across participants, and only divided participants into two groups who started with \tool or the baseline.

\subsection{Study Protocol}
We conducted the study through a video conferencing tool and in person, and each session took about 60--70 minutes.
At the beginning, we provided a short introduction to the study and obtained participants’ consent.
The participants then completed a pre-study survey to report their background.
We then introduced the study setting and task procedure and provided a tutorial for both interface conditions with a short demonstration of their features.
In this study, participants interacted with the interface via remote control.
Participants completed eight tasks, each with a 5-minute time limit.
For each task, we provided the task description and the agent's response, and we asked participants to submit the correct answer once they were confident.
They could decide whether to accept the agent's response or revise it if they found issues.
If needed, participants could use external search engines during the task.
For each task, we recorded task start and end times, participants’ final answers, and their interactions with the interface.
After each task, participants reported their confidence in the answer and the perceived effort using 7-point Likert-scale questions.
After completing all eight tasks, participants completed a post-study survey assessing their perceptions of the two interfaces and the usefulness of key interface features, both using 5-point Likert-scale questions.
At the end of the study, we conducted a semi-structured interview to collect participants’ solution strategies, feedback on both interfaces, and overall experience.
The study protocol was reviewed and determined to be exempt from IRB review under Exempt Category 3 (benign behavioral interventions with adult subjects).

\subsection{Data Collection}

We collected both quantitative and qualitative data during the study.
The complete survey materials are provided in Appendix~\ref{app:survey-instruments}.
Before the study, participants completed a recruitment form and a pre-study survey.
During the study, for each task, we recorded (1) completion time (from start and end timestamps), (2) answer correctness (correct/incorrect), and (3) self-reported confidence and effort ratings collected in the post-task survey.
After all tasks, participants completed a post-study survey that included overall interface preference, paired comparisons between \tool and the baseline, and feature-level usefulness ratings.
All sessions were screen-recorded, and participants’ interview responses were transcribed for qualitative analysis.

\subsection{Data Analysis}
To answer our three research questions, we conducted both quantitative and qualitative analyses of collected data.

For RQ1, we analyzed task completion time using a linear mixed-effects model and compared perceived effort using a paired Wilcoxon signed-rank test.
To further examine participants' verification behaviors, two researchers conducted coding of the screen recordings and compared the sequential patterns~\cite{jordan1995interaction}.

For RQ2, we analyzed verification accuracy using a logistic mixed-effects model, and further examined participants’ behaviors to understand the results.

For RQ3, we compared participants’ post-study ratings between conditions using paired Wilcoxon signed-rank tests and analyzed their use of individual \tool features from the coded behavioral traces.

For all RQs, we also conducted a thematic analysis~\cite{Clarke04052017} of the interview transcripts, which captured participants’ task walk-throughs and open-ended responses.
Two authors independently coded the transcripts to generate initial codes capturing the key information in participants' responses; these codes were consolidated and grouped into broader themes through discussion.
Differences in interpretation were resolved through discussion until consensus was reached.
Our thematic analysis resulted in 29 codes organized into six themes: verification strategy, use of evidence pages and highlights, trust and confidence calibration, error detection and correction, effort and workload, and interface design feedback.
Rather than reporting these themes separately, we integrate them with the quantitative and behavioral findings under the three research questions below.
We additionally analyzed confidence using a linear mixed-effects model with condition and task as fixed effects and participant as a random intercept.
\section{Results}

\begin{figure*}[t]
    \centering
    \begin{subfigure}[t]{0.49\linewidth}
        \centering
        \includegraphics[width=\linewidth]{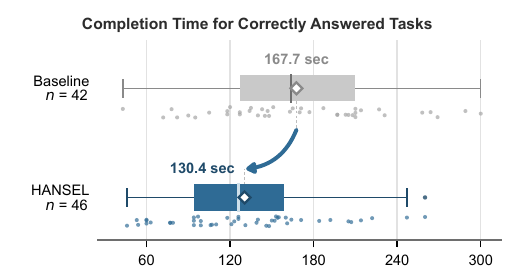}
        \caption{Completion time for correctly answered tasks ($n = 88$).}
        \label{fig:completion_time}
    \end{subfigure}
    \hfill
    \begin{subfigure}[t]{0.49\linewidth}
        \centering
        \includegraphics[width=\linewidth]{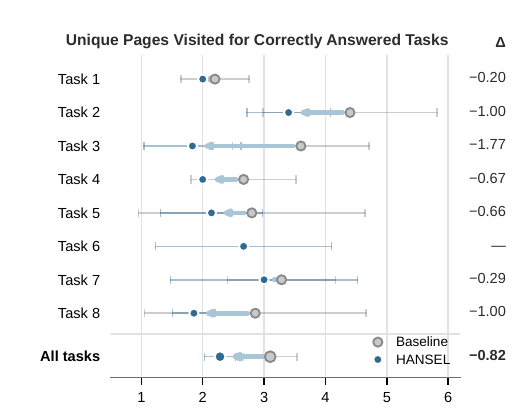}
        \caption{Unique pages visited per task for correctly answered tasks ($n = 88$). 
        $\Delta$ is \tool $-$ baseline in pages, so negative values mean fewer pages with \tool.}
        \label{fig:page_visited_per_task}
    \end{subfigure}
    \caption{Verification efficiency with \tool compared to the baseline.}
\end{figure*}

\subsection{RQ1: Verification Efficiency}
\label{subsec:rq1}

\subsubsection{Reduced Verification Time}

We first examined whether providing interactive evidence reduces the time and effort required for users to correctly verify web agents’ answers. 
We analyzed task completion time for trials in which participants submitted a correct answer ($N = 88$), so that completion time reflects time to a successfully verified answer. 

\Cref{fig:completion_time} summarizes the results.
A linear mixed-effects model with condition and task as fixed effects and participant as a random intercept showed significantly shorter completion times with \tool ($\mu = 130.4$ sec, $\sigma = 53.2$ sec) than with the baseline ($\mu = 167.7$ sec, $\sigma = 61.0$ sec; $\beta = -43.20$, $SE = 10.03$, $t = -4.31$, $p < .001$).
In the time model, three tasks showed significant differences relative to Task 1 (Task 3: $p = .008$, Task 7: $p = .005$, Task 8: $p = .018$), indicating non-trivial between-task differences in completion time.
As a robustness check, we repeated the analysis on all 112 trials and observed a nearly identical condition effect ($\beta = -44.73$, $SE = 9.54$, $t = -4.69$, $p < .001$).

\subsubsection{Reduced Perceived Effort}
In addition to completion time, participants rated their perceived effort after each task on a 7-point Likert scale. For correctly answered trials ($N = 88$), we compared participant-level median ratings using a paired Wilcoxon signed-rank test. Participants reported significantly lower effort with \tool than with the baseline (average median = 2.89 vs. 4.39; $V = 7$, $p = .012$, $r = .67$).

\subsubsection{Verification Strategies with Interactive Evidence}
\label{subsubsec:rq1_strategies}

\begin{figure}[t]
    \centering
    \includegraphics[
      width=1.0\textwidth,
      trim=0 1.2cm 0 2cm,
      clip]{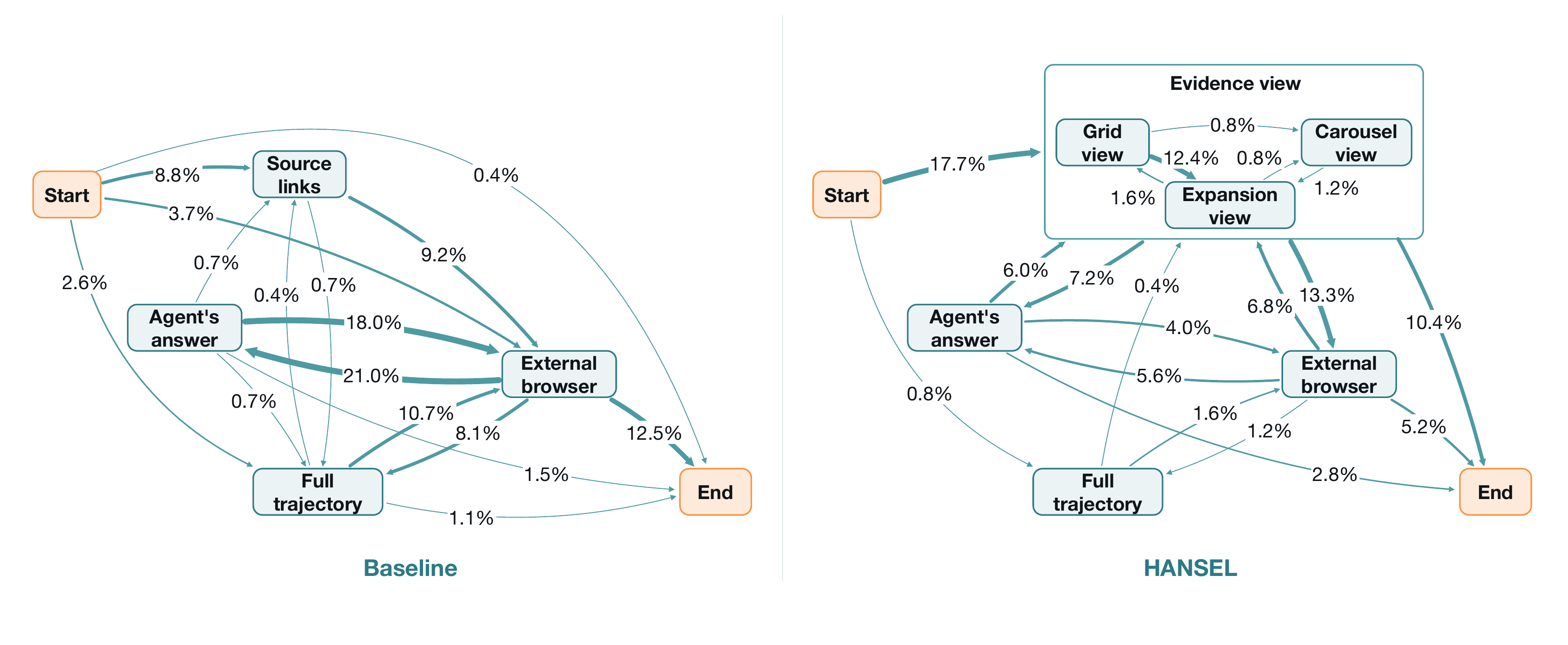}
    \caption{Verification pathways with \tool and the baseline.
    Nodes represent coded interaction states, with Grid, Carousel, and Expansion views grouped within \tool.
    Directed edges indicate transitions between states; edge labels show the percentage of all recorded transitions within each condition, aggregated across correctly answered trials (46 with \tool, 42 with the baseline). Percentages use the total number of transitions within the corresponding condition as the denominator. Table~\ref{tab:transition_matrix} reports the complete transition counts.}
    \label{fig:interaction_state_transition}
\end{figure}

To further understand the observed differences in verification time and effort, we compared participants’ verification behaviors across correctly answered trials ($N = 88$).
Two researchers coded the screen recordings into a set of interaction states capturing their interactions with the agent’s answer, full trajectory, source links, \tool evidence views (\eg grid view, carousel view, and expansion view), and external browser.
For example, a sequence of $\text{Start} \rightarrow \text{Source Link} \rightarrow \text{External Browser} \rightarrow \text{Agent’s Answer} \rightarrow \text{External Browser} \rightarrow \text{End}$ indicates that the participant first opened a source link, interacted with the corresponding webpage in an external browser, returned to inspect the agent’s answer, revisited the external browser, and then submitted their final judgment.
From these coded traces, we analyzed transitions between interaction states and recorded the number of distinct webpages (\ie unique URLs) visited during each task, including webpages embedded in \tool and those opened in an external browser.

Our results show that participants visited fewer web pages with \tool than with the baseline (Figure~\ref{fig:page_visited_per_task}) and followed different verification pathways across the two interfaces (Figure~\ref{fig:interaction_state_transition}).
In the baseline, participants frequently cross-checked the agent’s answer against information they found in the external browser. 
Transitions between answer inspection and the external browser accounted for 39.0\% of all transitions, compared with 9.6\% with \tool. 
Participants also moved back and forth between the full trajectory and the external browser, with these bidirectional transitions accounting for 18.8\% of all transitions. 
In these cases, participants used the trajectory to follow the agent’s actions and manually reproduce them in the external browser to verify the answer. 
For example, in Task 5, P13 repeatedly returned to the trajectory between visits to external webpages, reproducing the agent’s sorting and filtering steps one by one. 
In 23.8\% of transitions from the start (3.7\% of all transitions), participants went directly from the start of the task to the external browser, re-executing the task independently without relying on the agent’s answer.
This helps explain how participants reached correct answers in the baseline while requiring greater effort.

With \tool, participants often started verification with the evidence views.
Among transitions from the start, 95.7\% went directly to an evidence page (17.7\% of all transitions), compared with 4.3\% to the full trajectory (0.8\% of all transitions) and 0.0\% to the external browser. 
Beyond inspecting the evidence, participants could also interact directly with the evidence pages and sometimes reached their final judgment from them.
Among transitions to the end, 56.5\% came directly from the evidence views (10.4\% of all transitions), compared with 28.3\% from the external browser (5.2\% of all transitions).
Although external browsing remained part of verification with \tool, it primarily complements inspection within the evidence views, with participants moving from evidence views to the external browser (13.3\%) and returning to the evidence views (6.8\%).
In 50.0\% of \tool trials (23/46), participants completed verification without leaving the interface, compared with only 4.8\% (2/42) of baseline trials.

Together, these results show that with interactive evidence, participants could verify answers directly and reached correct answers faster and with less effort.

\subsection{RQ2: Verification Accuracy}

\subsubsection{Accuracy Patterns Across Verification Conditions}
From RQ1, we confirmed that interactive, evidence-based post-hoc verification can be more efficient.
However, while \tool’s evidence extraction achieved high precision, it may still miss important evidence, confusing the participants, and may lead users to rely heavily on the selected evidence even when the agent’s answer is incorrect.
To examine whether interactive evidence does not impose overreliance or confusion to users, we fit a logistic mixed-effects model predicting submission accuracy, with condition and task as fixed effects and participant as a random intercept.
We included task as a fixed effect to control for inherent differences in difficulty across the eight tasks.
Participant was modeled as a random intercept to account for individual differences, as observations from the same participant are not independent in our within-subjects design.

The mixed-effects model found no significant difference between conditions ($\beta = 0.71$, $SE = 0.67$, $z = 1.06$, $p = .29$).
Although the difference was not statistically significant, we found that the accuracy level was actually higher with \tool, at 82.1\% when the baseline accuracy was 75.0\%.
The difference was consistent for both cases when the shown agent's answer was correct and incorrect:
When the agent's answer was correct, participants submitted correct answers in 89.3\% of baseline trials and 96.4\% of \tool trials, and when the agent's answer was incorrect, accuracy was 60.7\% with the baseline and 67.9\% with \tool.

\subsection{RQ3: Usefulness of \tool}

\begin{figure}[t]
    \centering
    \includegraphics[width=\linewidth]{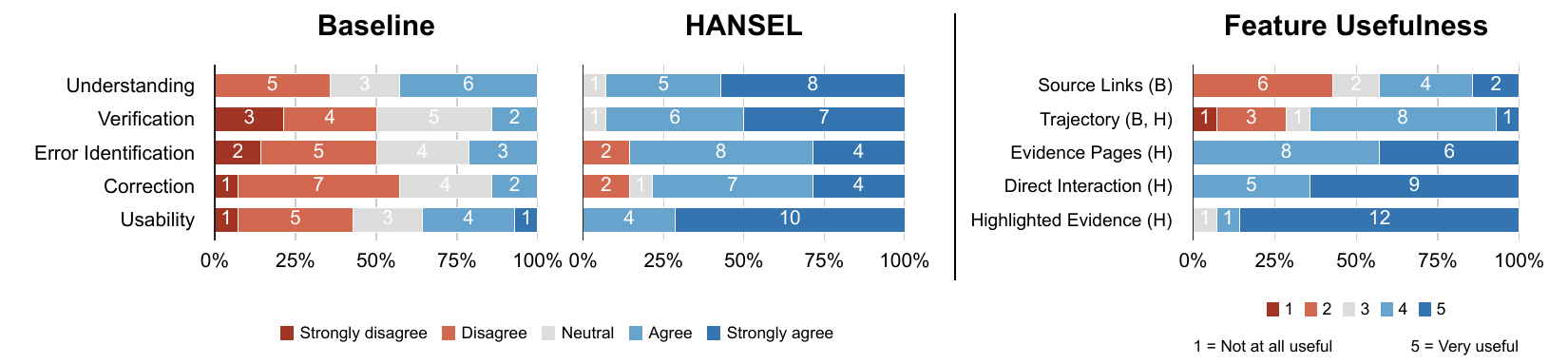}
    \caption{(Left) Summary of responses to the post-study survey comparing the baseline and \tool across five questions. (Right) Distribution of participant ratings for the usefulness of individual interface features.}
    \Description{Two survey charts: the left compares baseline and HANSEL ratings for understanding, verification, error identification, correction, and usability; the right shows usefulness ratings for individual interface features.}
    \label{fig:poststudy_hansel_vs_baseline}
\end{figure}

\subsubsection{Perceived Usefulness of \tool}
To answer RQ3, we analyzed the post-study survey results.
Participants rated the two interfaces across five dimensions: \textit{Understanding} (``helped me understand how the agent arrived at the answer''), \textit{Verification} (``made it easy to verify the agent’s response''), \textit{Error Identification} (``helped me identify potential problems in the agent’s response''), \textit{Correction} (``helped me revise the response when it was wrong''), and \textit{Usability} (``easy to learn and use'').
We used participant-level paired, two-sided Wilcoxon signed-rank tests to compare the 5-point Likert ratings for each dimension.

Figure~\ref{fig:poststudy_hansel_vs_baseline} summarizes the results.
Participants consistently rated \tool higher than the baseline across all five dimensions. 
Participants rated \tool to be significantly easier to use ($\mu=2.93 \rightarrow 4.71$, $p = .0045$) and to verify the agent’s response ($\mu=2.43 \rightarrow 4.43$, $p = .0013$). 
Using \tool, participants found it significantly easier to understand how the agent arrived at the answer ($3.07 \rightarrow 4.50$, $p = .0017$) and revise the response when it was wrong ($2.50 \rightarrow 3.93$, $p = .0045$). 
Participants also rated \tool higher for perceived support in identifying potential problems in the agent’s answer ($2.57 \rightarrow 4.00$, $p = .0127$).

Participants also rated individual interactive evidence features highly, such as \textit{Highlighted Evidence} (4.79/5), \textit{Direct Interaction with Pages} (4.64/5), and \textit{Evidence Pages with Titles} (4.43/5), with 93\%, 100\%, and 100\% of responses at 4 or above, respectively. 
In contrast, participants gave comparatively lower ratings to \textit{Source Links} (3.14/5; 43\% $\geq$ 4) and \textit{Step-by-step Trajectory} (3.36/5; 64\% $\geq$ 4) (Figure~\ref{fig:poststudy_hansel_vs_baseline}).

\subsubsection{Use of Interactive Evidence Features.}
To complement the perceived usefulness, which could be susceptible to social desirability bias~\cite{grimm_social_2010}, we also examined how participants actually used the individual features by reviewing screen recordings and analyzing the interview transcripts of the answers to the question, ``Which features of the interface did you find useful or not useful?''.

\paragraph{Direct Interaction with Evidence Pages.}
Overall, evidence pages were the most broadly used feature in \tool.
All participants assigned to \tool used evidence pages in all 56 \tool trials, and in 54 of them, they did so immediately after reading the agent’s answer.

Among the 46 correctly completed \tool trials, participants completed verification entirely within the interface in 23 trials: 19 using the grid view and 4 using the carousel view.
The reconstructed evidence pages provided task-relevant information and allowed participants to continue verification from the agent’s intermediate results without starting from scratch. 
In 10 of the remaining trials, participants moved from an evidence page to the browser, and in the other 13, they moved between the evidence pages and the browser to further check or confirm the information.
When we asked participants why they expanded evidence pages in an external browser, six participants said the embedded page was too small to read, and two attributed this to the Zoom remote-control setup used in our study.
\textit{``I opened it because\ldots I see the screen in the Zoom, so it's too small for me, so that's why I opened up.''} (P4)
While many participants had to expand the in-app interactive page in an external browser, it still indicates that evidence pages in \tool are useful as an entry point for verification.
\textit{``It already showed me the two pages of the evidence, so I just need to [expand] it to the bigger website to check the price.''} (P8)

\paragraph{Grid View vs. Carousel View}
Participants had different preferences for the two evidence layouts: the grid view and the carousel view.
Eight participants used only the grid view across all four \tool tasks, while P7, P11, and P13 relied primarily on the carousel view; P11 and P13 never used the grid view and consistently navigated through the carousel page by page.
In the grid view, participants often used the expand button to expand individual pages within \tool and inspect the highlighted evidence in more detail: 81\% of grid visits were followed by an expansion.
\textit{``First I checked the title, because it kind of helped me, like, which step I should check carefully.''} (P4)
Some participants preferred the carousel view because it supported sequential navigation without additional interactions.
\textit{``Between these two, I definitely prefer the carousel, since, it's like, in order for review.''} (P7) 
This suggests that verification strategies vary by person, showing the need to support diverse mechanisms.

\paragraph{Highlighted Evidence.}
In our interviews, ten participants reported that highlights helped direct their attention to relevant evidence, particularly on unfamiliar websites.
This guidance helped them identify where to focus and perform key verification actions more quickly.
\textit{``The red highlighting part is just, like, very useful to figure out, like, what on earth is it pointing at.''} (P7)
\textit{``It directs your attention to the most relevant places, then you immediately know how the pricing structure is like.''} (P9)
One participant whose first language was not English mentioned that the highlights reduced the effort of reading through the agent’s prose by directing attention to the most relevant information on the page.
\textit{``English is not my first language, so if the agent shows the pages [with] a lot of sentences without any highlights, it will cost me more time to understand or translate the English into Chinese.''} (P8)
Participants also provided feedback on highlights.
P3 noted that the highlights sometimes blended with the styling of the underlying webpage (\textit{``The Reddit page is, like, mostly red, and so when you have a highlight of red, it doesn't\ldots really pop out.''}).
Future designs could adapt highlight styles to different webpages or incorporate additional visual cues such as tooltips.

\section{Discussion} 
\label{sec:discussion}

We discuss the implications of our findings and the limitations of the current work.

\subsection{Implications}
\label{subsec:implication}

\subsubsection{From Passive Inspection to Interactive Verification.}
Our findings show that interactive, evidence-based post-hoc verification can make agent verification more efficient.
Participants correctly verified agent answers significantly faster and with less perceived effort using \tool than with the baseline.
Their interaction patterns further explain why: participants actively engaged with the extracted evidence and could inspect or correct errors directly within \tool, whereas with the baseline, they often moved between the interface and external webpages to understand and re-execute the agent’s actions.
This pattern echoes prior work on cognitive forcing functions in AI-assisted decision-making~\cite{buccinca2021totrustortothink} and extends this perspective to agent verification.
Rather than leaving users to passively inspect this information, verification interfaces for AI agents should provide natural entry points for users to inspect, interact with, and act on the agent’s work.
Future work could explore how to design such entry points for more complex agent executions, where both the execution trace and the agent’s answer may be substantially longer.
In such cases, evidence may need to be organized beyond the current flat grid and carousel views to help users identify which parts of the agent’s work warrant closer inspection.
Such support should also preserve the interactive nature of the evidence, allowing users to directly inspect and act on the relevant execution states.

\subsubsection{Critical Engagement with Evidence.}
After each task, we collected participants’ self-reported confidence in their answer.
Participants reported significantly higher confidence with \tool than with the baseline ($\mu = 6.05$ vs. $4.82$ on a 7-point scale; $\beta = 1.23$, $SE = 0.25$, $t = 4.92$, $p < .001$).
While confidence can be desirable in AI-assisted decision-making~\cite{sure2024ma}, our qualitative findings suggest a potential risk of over-trusting the presented evidence.
Some participants treated evidence pages as a starting point for independent verification, while others primarily checked whether the agent had taken the expected actions and accepted the answer without carefully examining whether those actions were correct.
While this tendency did not result in lower overall verification accuracy, as supported by our RQ2 results (82.1\% with \tool vs. 75.0\% with the baseline, with no significant difference between conditions), these observations still highlight the need to better support critical engagement with evidence.
To help users pay closer attention to the evidence, future verification interfaces could, for example, flag error-prone claim types or uncertain steps, or incorporate cognitive forcing functions that prompt users to inspect specific parts of the evidence before accepting an answer.
More broadly, future work should explore how to encourage independent verification when presenting supporting information from agent executions.

\subsubsection{Beyond Tasks with Bounded Answers.}
While our evaluation focused on tasks with clear ground-truth answers to ensure we can conduct user studies in a consistent and practical manner, there were many other interesting error types and verification needs that we believe \tool can support further.
One noteworthy case occurs when agents provide answers without taking the appropriate actions.
For example, in one task we examined (\textit{``Which bar is closest to Mummers Museum in Philadelphia and is wheelchair accessible?''}), the agent mentioned \textit{``Answering using domain knowledge''} in its reasoning rather than grounding the answer in information from a visited webpage.
While this could still be correct, it is neither verifiable nor trustworthy, given the known concerns around LLM hallucination.
In this case, surfacing the evidence pages, in this case 0, can make such gaps easily visible, helping users recognize the need for further verification.
Similarly, for longer-horizon tasks where the user can expect a larger evidence set, it will be easy to identify suspicious agent outputs when only a few pieces of evidence are shown to the user.
For tasks where correctness is subjective rather than verifiable against ground truth, what constitutes useful evidence may change.
In preference-driven tasks such as restaurant recommendation, for example, evidence could capture the decision points and alternatives considered by the agent, allowing users to assess whether its choices reflect their preferences and revise them when needed.
More broadly, our findings suggest that evidence should be adapted to the structure of the task. 
Future agent systems could present such task-relevant evidence as interactive artifacts alongside agent outputs, enabling users to inspect, question, and revise the agent’s decisions.

\subsection{Limitations}
\label{subsec:limitation}

Although \tool achieved strong evidence extraction performance across two web agent benchmarks, our evaluation covered a limited set of tasks and agent trajectories, which may limit the generalizability of our findings. 
While AssistantBench~\cite{yoran2024assistantbenchwebagentssolve} and Online-Mind2Web~\cite{xue2025illusionprogressassessingcurrent} are designed to reflect realistic web tasks, future work should evaluate and improve the extraction algorithm across a broader range of agents, websites, and task types.

Our controlled user study has several limitations.
While we made efforts to match task pairs by type and complexity, individual tasks may vary in difficulty, and the fixed task order may introduce learning effects.
The study setting also differs from everyday use of agent systems. 
Participants interacted with both interfaces through Zoom's remote control feature, which may have introduced latency and reduced visual quality. 
The researcher was present throughout each session, and participants were explicitly asked to submit correct answers and were aware that their behavior was being observed, which may have encouraged more verification than they would perform in everyday use.
Our single-session study also does not capture how verification practices may change as users gain experience with an agent or with \tool over time.
Future research could examine how users engage in verification in longer-term, real-world settings.

\section{Conclusion}
We designed \tool, a system that presents evidence from web agent trajectories as interactive, verifiable artifacts.
Rather than requiring users to passively read through overwhelming logs or trust LLM-generated summaries, \tool surfaces a compact set of evidence pages that contribute to the agent's answer and presents them as live, navigable views with preserved page state.
Our technical evaluation showed accurate evidence extraction (F1 = 0.831) with a 54.3\% reduction in trajectory volume.
A user study with 14 participants demonstrated that \tool enabled more efficient verification while improving usability. 
We hope that \tool's approach of surfacing interactive evidence for agent verification can inspire further research and adoption in real-world agent products, where robust verification support remains an open and pressing need.

\bibliographystyle{ACM-Reference-Format}
\bibliography{references}

\appendix

\section{Additional Details for Empirical Analysis}
\label{app:empirical_analysis_details}

\subsection{Annotation Procedure}

For each assigned task, annotators received the task, the agent's final output, and its full trajectory.
Each trajectory consists of several steps. Each step records the agent's action, the URL of the webpage on which it occurred, and the corresponding step content.
After reading the task and the agent's final output, annotators followed the trajectory step by step and determined whether each step was an evidence step.

\subsection{Annotation Protocol}

\paragraph{Evidence step.}
A step was labeled as an \emph{evidence step} if its content or visible page state was needed to trace or verify how the agent arrived at its final answer.
Evidence steps include both process evidence, such as a search query, selected filter, sorting criterion, or result page that reveals how the agent navigated toward the answer, and content evidence, such as a relevant fact, value, constraint, or result used in producing the answer.
When several consecutive operations formed a single interaction unit, such as selecting a filter through a dropdown, only the final step showing the resulting state was labeled as an evidence step.
When the final answer was simply reaching a specific page, only the final landing-page step was labeled as an evidence step.

\paragraph{Non-evidence step.}
A step was labeled as a \emph{non-evidence step} if any of the following held:

\begin{itemize}
    \item The page was inaccessible or had no usable content, such as a blocked page (CAPTCHA, login/paywall, or bot protection), a page that failed to load, a blank page, or an error page.
    \item The step belonged to a plan the agent later changed or abandoned, and the plan was not part of the path that led to the reported final answer.
    \item The step was action overhead, such as loading a homepage or landing page, performing a generic scroll or click, dismissing cookie or consent notices, or switching tabs.
    \item The step repeated information already shown by another step and added no new information needed for the answer.
    \item The step explored information that was not used in the final answer.
    \item The step was an intermediate operation whose resulting state was fully represented by a later step on the same page.
\end{itemize}

\paragraph{Evidence Page.}
A page was labeled as an \emph{evidence page} if it contained at least one evidence step (label = 1).
If a trajectory contained no evidence steps, it contained no evidence pages.
A page visit was defined as a consecutive sequence of steps sharing the same URL. Non-consecutive visits to the same URL were counted as separate page visits.
For example, in a trajectory \texttt{page1} $\rightarrow$ \texttt{page2} $\rightarrow$ \texttt{page3} $\rightarrow$ \texttt{page1}, the two \texttt{page1} visits were counted as separate pages.

\paragraph{Evidence Snippet.}
A snippet extracted by the LLM was judged accurate if both of the following held:

\begin{itemize}
    \item The snippet was observable on the website. If the page content had changed since the agent visited it, the snippet still counted if it was present in the agent's observation.
    \item The snippet directly captured content or visible page state relevant to the final answer, such as a search query, selected filter, a value, constraint, keyword, phrase, or paragraph.
\end{itemize}

\section{Prompts for LLM-Based Evidence Extraction}
\label{app:llm-evidence-prompt}

\begin{promptbox}[Prompts for LLM-Based Evidence Extraction]
Identify the minimal set of visited pages and steps needed to trace or verify how the agent arrived at its final answer.

Input:
The trajectory is given as a list of pages; each page is a run of consecutive steps sharing one URL.
<payload>

Copy `agent_answered`, `answer_text`, and `answer_step_num` exactly from `final_answer`.
Do not revise the answer or invent facts, URLs, step numbers, labels, values, or evidence.

Evidence does not need to establish that the final answer is correct.
Keep a step only if its content or visible page state is needed to trace or verify how the agent arrived at the final answer.

Prefer the latest step that fully shows the relevant state or evidence.
If a later kept step already shows the same information more completely, drop the earlier step.

For a multi-step search, filter, or form sequence:
- keep the final page or state that visibly contains the completed query, selections, or result;
- drop intermediate typing, clicking, adding, scrolling, and partial states when they are fully shown later;
- keep a search step only when its exact query, source choice, or retrieval path is not visible in a later kept step.

Always drop generic homepages, cookie actions, generic navigation, waiting, scrolling, failed or abandoned branches, blocked pages, and repeated information.

Use the shortest exact `evidence_snippet` from the trajectory:
- search: exact query;
- UI state: exact visible selected value or label;
- evidence: exact supporting text.
Use `[]` only if no exact text exists. Do not paraphrase snippets.

Return evidence pages in the order they were visited in the trajectory.
Within each page, list `evidence_snippets` in ascending `step_num` order.

If `agent_answered` is false, or if no visited step contains evidence, return an empty `evidence_pages`. 

Return JSON only:
{
  "agent_answered": <exact copy>,
  "answer_text": <exact copy>,
  "answer_step_num": <exact copy>,
  "evidence_pages": [
    {
      "url": "<copied URL>",
      "title": "<short page label>",
      "description": "<factual description>",
      "evidence_snippets": [
        {"step_num": <exact copy>, "text": "<exact text>"}
      ]
    }
  ]
}
\end{promptbox}
\section{User Study Details}
\label{app:survey-instruments}

This appendix records the questions and response formats used in the study surveys.

\subsection{Recruitment Survey}

The recruitment form asked the following questions.
\begin{enumerate}
    \item \textit{Email address}
    \item \textit{What is your gender identity?}
    \item \textit{Are you 18 years of age or older?}
    \item \textit{What is your occupation (field)? If you're a student, what course are you studying?} (free-text response).
    \item \textit{Have you ever used a conversational AI tool (e.g., ChatGPT, Claude, Gemini)?}
    \item \textit{Have you ever used the agent mode in these AI tools that can automatically browse websites and complete tasks?}
    \item \textit{Do you have any questions or additional comments?} (free-text response).
\end{enumerate}

\subsection{Pre-Study Survey}

\begin{enumerate}
    \item \textit{How good are you at completing tasks on websites (e.g., searching for products or finding places)?}
    \item \textit{How often do you use conversational AI tools (e.g., ChatGPT, Claude, Gemini)?}
    \item \textit{How often do you use the agent mode in these AI tools that can automatically browse websites and complete tasks?}
\end{enumerate}

\subsection{Task Response Form}

For each of the eight tasks below, participants entered an answer, rated their confidence with \textit{How confident are you in your answer?}, and rated effort with \textit{How much effort did you need to complete the task?} using a 7-point scale ranging from \textit{Very low} (1) to \textit{Very high} (7).

\begin{enumerate}
    \item \textbf{Task 1.} \textit{How much would I save by getting an annual membership for my family (2 adults, 1 kid age 5, 1 kid age 2) living in Washington State for the Seattle Children's Museum, compared to buying daily tickets, if we visit 4 times in a year?} 
    \item \textbf{Task 2.} \textit{On Flightaware, find discussions of the community and open one with the most replies.}
    \item \textbf{Task 3.} \textit{On BestBuy, search for NordicTrack with the lowest price.}
    \item \textbf{Task 4.} \textit{How much will I save by getting annual passes for a group of 4 adults and 1 student for the Philadelphia Museum of Art, compared to buying daily tickets, if we visit 5 times in a year (all non-consecutive days)?} 
    \item \textbf{Task 5.} \textit{On the Steam store, find the top game listed in the Steam Deck's top-played list over the past year. Then, count reviews from players who have played over 100 hours and primarily use a Steam Deck.}
    \item \textbf{Task 6.} \textit{On Apartments.com, search for the cheapest apartment in Detroit for a student.}
    \item \textbf{Task 7.} \textit{On Reddit, find the 12 Monkeys community and find the latest posts mentioning James Cole.}
    \item \textbf{Task 8.} \textit{On NFL.com, find the current NFL standings for the AFC East division and go to the page of the team that is in first place.}
\end{enumerate}

\subsection{Post-Study Survey}

\paragraph{Overall comparison.}
\begin{enumerate}
\item \textit{Which interface did you prefer overall?}
\item \textit{Which interface made it easier for you to verify whether the agent’s answer was correct?}
\end{enumerate}

\paragraph{Interface evaluation.}
Participants rated each of the following statements separately for \tool and the baseline on a 5-point scale from \textit{Strongly disagree} (1) to \textit{Strongly agree} (5).
\begin{enumerate}
\item The information provided by the interface helped me understand how the agent arrived at the answer.
\item The information provided by the interface made it easy to verify the agent’s response.
\item The information provided by the interface helped me identify potential problems in the agent’s response (e.g., using incorrect information or missing important steps).
\item When the agent’s response was wrong, the information provided by the interface helped me revise it.
\item The interface was easy to learn and use.
\end{enumerate}

\paragraph{Feature usefulness.}
Participants rated the usefulness of each feature (step-by-step trajectory, source links, evidence pages with short titles, direct interaction with evidence pages, and highlighted information) on a 5-point scale from \textit{Not at all useful} (1) to \textit{Very useful} (5).
\begin{enumerate}
\item Viewing the step-by-step trajectory in both interfaces (e.g., ``show 22 steps''). \item Viewing the source links in the source-link interface. \item Viewing evidence pages with short titles in the interactive-evidence interface. \item Interacting directly with the evidence pages (e.g., clicking and scrolling). \item Seeing highlighted information on the evidence pages (e.g., ``Global Gas Station'', ``4.3'').
\end{enumerate}

\subsection{Complete Transition Matrices}

\newcommand{\pct}[1]{\textcolor{gray}{\scriptsize(#1)}}
\begin{table*}[h]
\centering
\small
\setlength{\tabcolsep}{4pt}
\caption{Complete transition matrices for correctly answered trials ($N = 88$).
Each cell reports the number of directed transitions between interaction states and its percentage of all transitions within the corresponding condition (249 for \tool; 272 for the baseline).}
\label{tab:transition_matrix}
\begin{tabular}{@{}lrrrrrrrr@{}}
\toprule
\multicolumn{9}{@{}l}{\textbf{\tool} --- 46 trials, 249 transitions}\\[2pt]
\textit{from} $\downarrow$\,\textit{to} $\rightarrow$ & Answer & Traj. & Grid & Car. & Exp. & Ext. & End & \textit{total} \\
\midrule
\textsc{Start} & \textcolor{gray}{0} & 2\,\pct{0.8} & 28\,\pct{11.2} & 10\,\pct{4.0} & 6\,\pct{2.4} & \textcolor{gray}{0} & \textcolor{gray}{0} & 46\,\pct{18.5} \\
Agent's answer & -- & \textcolor{gray}{0} & 1\,\pct{0.4} & 5\,\pct{2.0} & 9\,\pct{3.6} & 10\,\pct{4.0} & 7\,\pct{2.8} & 32\,\pct{12.9} \\
Full trajectory & \textcolor{gray}{0} & -- & 1\,\pct{0.4} & \textcolor{gray}{0} & \textcolor{gray}{0} & 4\,\pct{1.6} & \textcolor{gray}{0} & 5\,\pct{2.0} \\
Grid view & 1\,\pct{0.4} & \textcolor{gray}{0} & -- & 2\,\pct{0.8} & 31\,\pct{12.4} & \textcolor{gray}{0} & 1\,\pct{0.4} & 35\,\pct{14.1} \\
Carousel view & 6\,\pct{2.4} & \textcolor{gray}{0} & \textcolor{gray}{0} & -- & 3\,\pct{1.2} & 14\,\pct{5.6} & 6\,\pct{2.4} & 29\,\pct{11.6} \\
Expansion view & 11\,\pct{4.4} & \textcolor{gray}{0} & 4\,\pct{1.6} & 2\,\pct{0.8} & -- & 19\,\pct{7.6} & 19\,\pct{7.6} & 55\,\pct{22.1} \\
External browser & 14\,\pct{5.6} & 3\,\pct{1.2} & 1\,\pct{0.4} & 10\,\pct{4.0} & 6\,\pct{2.4} & -- & 13\,\pct{5.2} & 47\,\pct{18.9} \\
\midrule
\textit{total} & 32\,\pct{12.9} & 5\,\pct{2.0} & 35\,\pct{14.1} & 29\,\pct{11.6} & 55\,\pct{22.1} & 47\,\pct{18.9} & 46\,\pct{18.5} & 249\,\pct{100.0} \\
\bottomrule
\end{tabular}

\vspace{9pt}

\begin{tabular}{@{}lrrrrrr@{}}
\toprule
\multicolumn{7}{@{}l}{\textbf{Baseline} --- 42 trials, 272 transitions}\\[2pt]
\textit{from} $\downarrow$\,\textit{to} $\rightarrow$ & Answer & Traj. & Source & Ext. & End & \textit{total} \\
\midrule
\textsc{Start} & \textcolor{gray}{0} & 7\,\pct{2.6} & 24\,\pct{8.8} & 10\,\pct{3.7} & 1\,\pct{0.4} & 42\,\pct{15.4} \\
Agent's answer & -- & 2\,\pct{0.7} & 2\,\pct{0.7} & 49\,\pct{18.0} & 4\,\pct{1.5} & 57\,\pct{21.0} \\
Full trajectory & \textcolor{gray}{0} & -- & 1\,\pct{0.4} & 29\,\pct{10.7} & 3\,\pct{1.1} & 33\,\pct{12.1} \\
Source links & \textcolor{gray}{0} & 2\,\pct{0.7} & -- & 25\,\pct{9.2} & \textcolor{gray}{0} & 27\,\pct{9.9} \\
External browser & 57\,\pct{21.0} & 22\,\pct{8.1} & \textcolor{gray}{0} & -- & 34\,\pct{12.5} & 113\,\pct{41.5} \\
\midrule
\textit{total} & 57\,\pct{21.0} & 33\,\pct{12.1} & 27\,\pct{9.9} & 113\,\pct{41.5} & 42\,\pct{15.4} & 272\,\pct{100.0} \\
\bottomrule
\end{tabular}
\end{table*}

\end{document}